\documentstyle[aps,eqsecnum]{revtex}
 
\begin{document}
\title{Spin-spin effects in radiating compact binaries}
\author{L\'{a}szl\'{o} \'{A}. Gergely}
\address{Laboratoire de Physique Th\'{e}orique,
Universit\'{e} Louis Pasteur, 3-5 rue de l'Universit\'{e} 67084,
Strasbourg, France\\
and KFKI Research Institute for Particle and Nuclear Physics, Budapest 114,
P.O.Box 49, H-1525 Hungary}
\maketitle
 
\begin{abstract}
The dynamics of a binary system with two spinning components on an eccentric
orbit is studied, with the inclusion of the spin-spin interaction terms
appearing at the second post-Newtonian order. A generalized true anomaly
parametrization properly describes the radial component of the motion. The
average over one radial period of the magnitude of the orbital angular
momentum $\bar{L}$ is found to have no nonradiative secular change. All
spin-spin terms in the secular radiative loss of the energy and magnitude of
orbital angular momentum are given in terms of $\bar{L}$ and other constants
of the motion. Among them, self-interaction spin effects are found,
representing the second post-Newtonian correction to the $3/2$
post-Newtonian order Lense-Thirring approximation.
\end{abstract}
 
\section{Introduction}
 
Neutron-star$-$black-hole and black-hole$-$black-hole binaries are among the
most promising sources for the earth-based gravitational wave observatories,
such as the Laser Interferometric Gravitational Wave Observatory LIGO\cite
{LIGO}, VIRGO\cite{VIRGO}, GEO\cite{GEO} and TAMA\cite{TAMA}. They are also
important sources for the forthcoming Laser Interferometer Space Antenna
(LISA)\cite{LISA,LISA1}. It is generally agreed that in a post-Newtonian
description of the motion and of the gravitational radiation of these binary
systems a precision of $7/2$ post-Newtonian orders has to be reached. At
such high orders, due to the nonlinear character of the Einstein equations,
various multipolar contributions to the gravitational radiation are present.
 
Computations based on the general expectation that the orbit of a binary
will circularize have reached high post-Newtonian orders. However in various
physical situations, as described in Refs. \cite{GI}-\cite{HiBe}, the
eccentricity of the orbits can play a significant role. An equally important
feature of the binary system, often neglected, is the effect of the
individual spins on the motion and on the radiation of the binary. The
dynamics under the influence of the spins was investigated in Refs. \cite{BB}
-\cite{Kidder}. Spin contributions already appear at the $3/2$
post-Newtonian order, and exceptionally at the first post-Newtonian order
when angular evolutions are monitored. A complete description should allow
for both spin effects and eccentric motions.
 
In a recent series of papers Gergely, Perj\'{e}s, and Vas\'{u}th have
investigated the influence of the spins in the secular evolution and
radiation back reaction of binary systems on eccentric orbit up to $3/2$
post-Newtonian order. In the last paper \cite{GPV3} of these series, the
secular radiative evolution of the energy $E$ and magnitude of the orbital
angular momentum $L$ were given. Both $E$ and $L$ are constants of the
non-radiative motion up to $3/2$ post-Newtonian order. At this order, among
the spin effects, only the spin-orbit interaction is involved in the
radiative change of $E$ and $L$. The results were in perfect agreement with
computations of Rieth and Sch\"{a}fer \cite{RS} in a different, noncovariant
spin supplementary condition, with the energy and orbital momentum derived
from a different action.
 
Also in \cite{GPV3}, both the nonradiative and radiative evolution of three
angles\footnote{
An overhat denotes a unit vector.} $\kappa _{i}=\cos ^{-1}({\bf \hat{S}
_{i}\cdot \hat{L})},\ (i=1,2)$ and $\gamma =\cos ^{-1}({\bf \hat{S}_{1}\cdot
\hat{S}_{2})},$ characterizing the relative orientation of the spin vectors
{\bf \ }${\bf S}_{{\bf i}}$ and orbital angular momentum ${\bf L,}$ were
given. By considering both the spin-orbit and the spin-spin interactions,
the nonradiative change of the angles receives both first and $3/2$
post-Newtonian order contributions, while the radiative changes are of $3/2$
post-Newtonian order. Remarkably, the radiative change of the spins
themselves gave no secular contribution to the radiative evolution of the
angles.
 
Earlier results, pertinent for the physical situations of spinning binaries
on eccentric orbit with comparable masses, but one of the spins dominating
over the second \cite{GPV2}, and a nonspinning particle orbiting about a
spinning object (the Lense-Thirring approximation) \cite{GPV1} (see also
Ref. \cite{Ryan}), can be obtained as limiting cases of Ref. \cite{GPV3}.
 
The computations in Refs. \cite{GPV3}, \cite{GPV2}, and \cite{GPV1} were
facilitated by the solution of the radial motion of the binary in terms of a
generalized true anomaly parameter. Then all relevant integrals presented
the remarkable property of having the only pole in the origin. Later a
systematic analysis \cite{param} has yield the same type of parametrization
for a wide class of perturbations of the Keplerian motion, including the
generic perturbing force of Brumberg \cite{Brumberg,Soffel}.
 
The main impediment in generalizing the description of Ref. \cite{GPV3} to
include all spin-spin effects was that at the second post-Newtonian order
the magnitude $L$ of the orbital angular momentum, even in the absence of
the radiation, is not a conserved quantity. This feature of the spin-spin
type perturbation of the Keplerian motion renders the problem even outside
the general framework settled in Ref. \cite{param}.
 
The purpose of this paper is to overcome the above difficulty and to give a
solution to the radial motion in terms of the generalized true anomaly
parametrization, which is valid at the second post-Newtonian order in the
presence of the spin-spin interaction terms. This task will be completed by
a careful analysis of the specific form of the evolution of the orbital
angular momentum. In Sec. II we study the radial motion. We compute the
instantaneous expression of $L$ and then we find that it has {\it no secular
change}. Therefore the angular average value $\bar{L}$ has a deep
significance: it has no other change during the whole adiabatic regime of
the evolution of the system, then the radiative one. The average magnitude $
\bar{L}$ of the orbital angular momentum provides the missing constant of
the motion at the second post-Newtonian order, and in fact it coincides with
the previously employed $L$ up to $3/2$ post-Newtonian orders. As a
by-product, the radial equation of motion is derived.
 
Based on the analysis of the radial motion, in Sec. III we introduce the
generalized true and eccentric anomaly parametrizations. The latter is
employed in the computation of the radial period, which is found to have the
Keplerian expression in terms of the energy of the perturbed motion.
 
Section IV contains the main results of the paper: the spin-spin
contributions to the instantaneous and secular losses of the energy and
magnitude of orbital angular momentum. The computations are considerably
facilitated by the advantageous properties of the true anomaly
parametrization. Among these losses self-interaction type contributions
appear.
 
In a second post-Newtonian order description of the compact binary there are
no other spin terms in the radiative losses of energy and magnitude of
orbital angular momentum then the spin-spin terms computed in this paper,
together with the corresponding spin-orbit expressions from Ref. \cite{GPV3}
. We note here that it is equally possible to give a description in terms of
the time average $\langle L\rangle $ of the magnitude of orbital angular
momentum. However, as will be deduced in the Appendix, the relation between $
\langle L\rangle $ and $\bar{L}$ is cumbersome. The description in terms of $
\bar{L}$ turns out to be simpler. We keep the velocity of light $c$ and the
gravitational constant $G$ in all expressions.
 
\section{The radial motion}
 
The Keplerian motion of a reduced mass particle $\mu $ in the gravitational
potential generated by the total mass $m=m_{1}+m_{2}$ is governed by the
Lagrangian
 
\begin{equation}
{\cal L}_{N}=\frac{\mu {\bf v}^{2}}{2}+{\frac{Gm\mu }{r}\ ,}
\end{equation}
or equivalently, by the acceleration
\begin{equation}
{\bf a}_{N}=-\frac{Gm{\bf r}}{r^{3}}\ .  \label{aN}
\end{equation}
 
Among other dynamical quantities, the Newtonian energy $E_{N}$ and magnitude
$L_{N}$ of the Newtonian orbital angular momentum are conserved. In terms of
the spherical polar coordinates $r,\theta $ and $\varphi $ they are
expressed as
\begin{eqnarray}
E_{N} &=&{\frac{\mu v^{2}}{2}}-{\frac{Gm\mu }{r}}\   \nonumber \\
&=&\ {\frac{\mu }{2}[\dot{r}^{2}+r^{2}(\dot{\theta}^{2}+\sin ^{2}\theta \
\dot{\varphi}^{2})]}-{\frac{Gm\mu }{r}}\ ,  \label{EN} \\
L_{N}^{2} &=&{\mu ^{2}}r^{4}(\dot{\theta}^{2}+\sin ^{2}\theta \ \dot{\varphi}
^{2})\ .  \label{LN}
\end{eqnarray}
 
From the first expression of $E_{N}$ we find the velocity $v$. By suitably
combining the second expression for $E_{N}$ with Eq. (\ref{LN}) we obtain
the equation for the radial motion given by $\dot{r}$:
\begin{eqnarray}
\ v^{2} &=&\frac{2E_{N}}{\mu }+\frac{2Gm}{r}\ ,  \label{v2} \\
\dot{r}^{2} &=&\frac{2E_{N}}{\mu }+\frac{2Gm}{r}-\frac{L_{N}^{2}}{\mu
^{2}r^{2}}\ .  \label{rdot2}
\end{eqnarray}
 
When the Keplerian motion is perturbed, Eqs. (\ref{v2}) and (\ref{rdot2})
still hold as identities. However the Newtonian expressions $E_{N}$ and $
L_{N}$ are not constants of motion any more. The procedure used in Refs.
\cite{GPV3}, \cite{GPV2}, and \cite{GPV1} was to express them in terms of the
constants of motion $E,L,\kappa _{i},$ and $\gamma $ of the perturbed motion
and of the radius $r$ alone. There is no simple way to fulfill a similar
task when the perturbation is due to spin-spin interaction, as will be seen
in what follows. In fact a milder requirement will be enforced: $v^{2}$ and $
\dot{r}^{2}$ will be expressed in terms of constants of motion and a
suitably introduced true anomaly parameter $\chi $.
 
The motion of the binary under the influence of the spin-spin interactions
is governed by the Lagrangian \cite{KWW}
 
\begin{eqnarray}
{\cal L} &=&{\cal L}_{N}+{\cal L}_{SS}\ ,  \nonumber \\
{\cal L}_{SS} &=&\frac{G}{c^{2}r^{3}}\left[ \left( {\bf S}_{{\bf 1}}\cdot
{\bf S}_{{\bf 2}}\right) -\frac{3}{r^{2}}\left( {\bf r\cdot S}_{{\bf 1}
}\right) \left( {\bf r\cdot S}_{{\bf 2}}\right) \right] \ .  \label{LagSS}
\end{eqnarray}
We have deliberately omitted those first and second post-Newtonian order
terms (PN, 2PN) which do not contain the spins and the spin-orbit (SO)
terms. Their contributions were discussed previously in Refs. \cite{GI} and
\cite{GPV3}, and they add linearly to the spin-spin perturbative terms
computed in this paper.
 
An equivalent characterization of the motion is provided by the acceleration
\begin{eqnarray}
{\bf a} &=&{\bf a}_{N}+{\bf a}_{SS}\ , \\
{\bf a}_{SS} &=&-\frac{3G}{c^{2}\mu r^{7}}\{r^{2}\left[ {\bf r}\left( {\bf S}
_{{\bf 1}}\cdot {\bf S}_{{\bf 2}}\right) +{\bf S}_{{\bf 1}}\left( {\bf r}
\cdot {\bf S}_{{\bf 2}}\right) +{\bf S}_{{\bf 2}}\left( {\bf r}\cdot {\bf S}
_{{\bf 1}}\right) \right]  \nonumber \\
&&-5{\bf r}\left( {\bf r}\cdot {\bf S}_{{\bf 1}}\right) \left( {\bf r}\cdot
{\bf S}_{{\bf 2}}\right) \}  \label{aSS}
\end{eqnarray}
The perturbed Keplerian motion (\ref{LagSS}) is characterized \cite{KWW},
\cite{Kidder} by the conservation of the total energy
 
\begin{equation}
E=E_{N}+E_{SS}\ ,\qquad E_{SS}=-{\cal L}_{SS}\   \label{ESS}
\end{equation}
and of the total angular momentum vector
\begin{equation}
{\bf J=L+S\ ,}  \label{J}
\end{equation}
where ${\bf L=L}_{N}=\mu \left( {\bf r\times v}\right) $ is the total
orbital angular momentum (there is no spin-spin contribution to the orbital
angular momentum and the PN, 2PN, SO contributions are not listed here). $
{\bf S=S}_{{\bf 1}}+{\bf S}_{{\bf 2}}$ is the total spin. Both spin vectors
undergo a precessional motion due to the spin-spin interaction
 
\begin{equation}
{\bf \dot{S}_{i}}={\frac{G}{c^{2}r^{3}}}\left[ \frac{3}{r^{2}}({\bf r\cdot
S_{j}}){\bf r-S_{j}}\right] \times {\bf S_{i}}\ ,  \label{Sprec}
\end{equation}
where $i\neq j.$
 
From the conservation of the total angular momentum ${\bf J}$, Eq. (\ref{J})
and the precession equations (\ref{Sprec}) we find that the orbital angular
momentum evolves as
\begin{equation}
{\bf \dot{L}=}\frac{3G}{c^{2}r^{5}}\left[ ({\bf r\cdot S}_{{\bf 2}}){\bf S}_{
{\bf 1}}+({\bf r\cdot S}_{{\bf 1}}){\bf S}_{{\bf 2}}\right] \times {\bf r\ .}
\label{Lvecdot}
\end{equation}
The magnitude of the orbital angular momentum changes according to $\dot{L}=
{\bf \hat{L}}\cdot {\bf \dot{L},}$ thus
 
\begin{eqnarray}
\dot{L}=-\frac{3G}{c^{2}r^{5}}S_{1}S_{2} &\Bigl\{&({\bf r\cdot \hat{S}}_{
{\bf 2}})\left[ {\bf \hat{L}\cdot (r\times \hat{S}}_{{\bf 1}})\right]
\nonumber \\
&+&({\bf r\cdot \hat{S}}_{{\bf 1}})\left[ {\bf \hat{L}\cdot (r\times \hat{S}}
_{{\bf 2}})\right] \Bigr\}\ .  \label{Ldot}
\end{eqnarray}
This change in $L$ is of second post-Newtonian order. Therefore Eq. (\ref
{Ldot}) can be simplified by inserting the previously derived Newtonian
expressions, Eqs. (2.19) and (2.22) of Ref. \cite{GPV3}:
 
\begin{eqnarray}
{\bf r\cdot \hat{S}_{i}} &=&r\sin \kappa _{i}\cos (\psi -\psi _{i})\ ,
\label{aux1} \\
{\bf \hat{L}\cdot ({r}\times \hat{S}_{i})} &=&-r\sin \kappa _{i}\sin (\psi
-\psi _{i})\ ,  \label{aux2}
\end{eqnarray}
where the angles $\psi ,\ \psi _{i}$ are subtended by the node line with the
position ${\bf r}$ and the projections of the spins in the plane of the
orbit, respectively, see Fig. 1 in Ref. \cite{GPV3}. In order to derive Eq. (
\ref{aux2}) the auxiliary relation
\begin{equation}
{\bf v\cdot \hat{S}_{i}}=\dot{r}\sin \kappa _{i}\cos (\psi -\psi _{i})-\frac{
L}{\mu r}\sin \kappa _{i}\sin (\psi -\psi _{i})  \label{aux3}
\end{equation}
was employed. The angles $\psi _{i}$ are given implicitly in terms of $
\kappa _{i}$ and $\gamma $ by Eq. (2.20) of Ref. \cite{GPV3}
 
\begin{equation}
S_{1}\sin \kappa _{1}\cos \psi _{1}+S_{2}\sin \kappa _{2}\cos \psi _{2}=0\
\label{ang1}
\end{equation}
and by the spherical cosine identity
 
\begin{equation}
\cos \gamma =\cos \kappa _{1}\cos \kappa _{2}+\sin \kappa _{1}\sin \kappa
_{2}\cos \Delta \psi ,  \label{ang2}
\end{equation}
where $\Delta \psi =\psi _{2}-\psi _{1}.$ We introduce the additional
notation $\bar{\psi}=(\psi _{1}+\psi _{2})/2.$
 
In terms of the above angles the variation of the magnitude of orbital
angular momentum takes the simple form
 
\begin{equation}
\dot{L}=\frac{3G}{c^{2}r^{3}}S_{1}S_{2}\sin \kappa _{1}\sin \kappa _{2}\sin
2(\psi -\bar{\psi})\ .  \label{LdotAng}
\end{equation}
We remark that none of the perturbations not treated here (PN, 2PN, SO)
would give contributions to the above expression for $\dot{L}.$ This is to
be contrasted with ${\bf \dot{L},}$ which receives also the SO-type
contribution [Eq. (2.11) in Ref. \cite{GPV3}].
 
The change in the magnitude of the orbital angular momentum in the second
post-Newtonian order is a feature not encountered in the discussion of the
PN, 2PN, SO effects. This was the major inconvenience in generalizing the
description of Ref. \cite{GPV3} to include the spin-spin contributions. We
will see in what follows how to handle this difficulty by means of an
alternative description based on a properly defined average value $\bar{L}$
of $L$.
 
The Newtonian limit of a generalized true anomaly parameter $\chi $ (to be
defined later in this paper) has to coincide with the Keplerian true
anomaly. This feature is enough to introduce it in second post-Newtonian
expressions already at this stage. As implied by Eq. (\ref{LdotAng}), the
magnitude $L(\chi )$ of the orbital angular momentum at some value of the
true anomaly $\chi \in [0,2\pi ]$ differs in the second post-Newtonian order
from $L_{0}=L\left( 0\right) $ :
\begin{equation}
L(\chi )=L_{0}+\int_{0}^{\chi }\dot{L}\frac{dt}{d\chi ^{^{\prime }}}d\chi
^{^{\prime }}\ .  \label{Lchi}
\end{equation}
We compute the integral by inserting Eq. (\ref{LdotAng}) and the following
Newtonian expressions:
\begin{eqnarray}
\frac{dt}{d\chi } &=&\frac{\mu r^{2}}{\bar{L}}\ ,  \label{dtdchiN} \\
r &=&\frac{\bar{L}^{2}}{\mu (Gm\mu +\bar{A}\cos \chi )}\ ,  \label{rN} \\
\psi  &=&\chi +\psi _{0}\ .  \label{psiN}
\end{eqnarray}
The angle $\psi _{0}$ is subtended by the direction of the periastron and
the node line, see Fig. 1 in Ref. \cite{GPV3} and we have denoted by $\bar{L}
$ the average value of $L(\chi )$ over the range $(0,2\pi ).$ The quantity $
\bar{A}$ defined by
\begin{equation}
\bar{A}=\left( G^{2}m^{2}\mu ^{2}+\frac{2E\bar{L}^{2}}{\mu }\right) ^{1/2}
\label{Abar}
\end{equation}
is the magnitude of the Laplace-Runge-Lenz vector for a Keplerian motion
characterized by $E$ and $\bar{L}$. The integration yields
 
\begin{eqnarray}
L(\chi ) &=&L_{0}+\frac{G\mu ^{2}}{2c^{2}\bar{L}^{3}}S_{1}S_{2}\sin \kappa
_{1}\sin \kappa _{2}\Theta \ ,  \nonumber \\
\Theta &=&3Gm\mu \left[ \cos 2(\psi _{0}\!-\!\bar{\psi})-\cos 2(\chi
\!+\!\psi _{0}\!-\!\bar{\psi})\right]  \nonumber \\
&+4&\bar{A}\left[ \cos 2(\psi _{0}\!-\!\bar{\psi})-\cos \chi \cos 2(\chi
\!+\!\psi _{0}\!-\!\bar{\psi})\right]  \nonumber \\
&&-2\bar{A}\sin \chi \sin 2(\chi \!+\!\psi _{0}\!-\!\bar{\psi})\ .
\label{LchiAng}
\end{eqnarray}
It is straightforward to verify that $L(0)=L(2\pi )=L_{0},$ because all
three terms of $\Theta $ independently vanish at $\chi =0,2\pi $. Therefore
{\it there is no change due to the spin-spin interaction in the magnitude of
the orbital angular momentum over one radial period}.
 
A second integration over the parameter $\chi $ from $0$ to $2\pi $ gives
the angular average of the magnitude of orbital angular momentum:
\begin{equation}
\bar{L}=\frac{1}{2\pi }\int_{0}^{2\pi }L(\chi )d\chi =L_{0}+\frac{G\mu
^{2}(3Gm\mu +4\bar{A})S_{1}S_{2}\beta _{0}}{2c^{2}\bar{L}^{3}}\ ,
\label{Lave}
\end{equation}
where we have introduced the notation
\begin{equation}
\beta _{0}=\sin \kappa _{1}\sin \kappa _{2}\cos 2(\psi _{0}-\bar{\psi})\ .
\label{beta0}
\end{equation}
In terms of $\bar{L}$ the expression of $L(\chi )$ reduces to
\begin{eqnarray}
L(\chi ) &=&\bar{L}-\frac{G\mu ^{2}}{2c^{2}\bar{L}^{3}}S_{1}S_{2}\sin
\!\kappa _{1}\sin \!\kappa _{2}\{2\bar{A}\cos [\chi \!+2(\!\psi _{0}\!-\!
\bar{\psi})]  \nonumber \\
&&+(3Gm\mu \!+\!2\bar{A}\cos \!\chi )\cos 2(\chi \!+\!\psi _{0}\!-\!\bar{\psi
})\}\ .  \label{LchiAng1}
\end{eqnarray}
 
A considerably simpler task is to find the spin-spin part $E_{SS}$ of the
energy. This can be expressed from Eq. (\ref{ESS}) by use of Eqs. (\ref{aux1}
) and (\ref{ang2}) [the azimuthal angle $\psi $ being eliminated by its
expression (\ref{psiN})]:
\begin{eqnarray}
E_{SS}(r,\chi ) &=&-\frac{G}{2c^{2}r^{3}}S_{1}S_{2}\alpha (\chi )\ ,
\label{ESSAng} \\
\alpha (\chi ) &=&3\cos \kappa _{1}\cos \kappa _{2}-\cos \gamma   \nonumber
\\
&-&3\sin \kappa _{1}\sin \kappa _{2}\cos 2(\chi \!+\!\psi _{0}\!-\!\bar{\psi}
)\ .  \label{alpha}
\end{eqnarray}
$E_{SS}$ is given solely in terms of constants of the motion, radius $r$ and
true anomaly parameter $\chi .$ Now we are in position to write the
expressions for $v^{2}$, Eq. (\ref{v2}) and for $\dot{r}^{2}$, Eq. (\ref
{rdot2}) in suitable form:
\begin{eqnarray}
\ v^{2} &=&\frac{2[E-E_{SS}(r,\chi )]}{\mu }+\frac{2Gm}{r}\ ,  \label{v2Ang}
\\
\dot{r}^{2} &=&\frac{2[E-E_{SS}(r,\chi )]}{\mu }+\frac{2Gm}{r}-\frac{L(\chi
)^{2}}{\mu ^{2}r^{2}}\ ,  \label{rdot2Ang}
\end{eqnarray}
with $L(\chi )$ and $E_{SS}(r,\chi )$ given by Eqs. (\ref{LchiAng1}) and (
\ref{ESSAng}).
 
As the angles $\kappa _{i}$ are constant up to the first post-Newtonian
order and $\gamma $ up to one-half post-Newtonian order [see Eqs. (2.17) in
Ref. \cite{GPV3}], they do not vary in the above expressions. In principle
it is possible to rewrite Eqs. (\ref{v2Ang}) and (\ref{rdot2Ang}) in terms
of constants of motion and $r$ alone. This is because the true anomaly
parameter $\chi $ enters only at Newtonian order, in which order the
relation $\chi =\chi (r)$ is given by Eq. (\ref{rN}). We avoid to write such
cumbersome expressions, as our task will be exactly the converse: to
parametrize all expressions in terms of $\chi $.
 
\section{The true and eccentric anomaly parametrizations}
 
In this section first we introduce the generalized true anomaly
parametrization $r=r(\chi )$ for the radial motion (\ref{rdot2Ang}). For
this purpose we have to find the turning points $r_{{}_{{}_{min}^{max}}}$
defined by
\begin{eqnarray}
r_{min} &=&r(0)\ ,\qquad \dot{r}^{2}(0)=0\ ,  \nonumber \\
r_{max} &=&r(\pi )\ ,\qquad \dot{r}^{2}(\pi )=0\ .  \label{turningdef}
\end{eqnarray}
As from Eq. (\ref{LchiAng1}) the magnitude of the orbital angular momentum
in the turning points is found in the form $L(0)=\bar{L}+\delta L_{-}$ and $
L(\pi )=\bar{L} +\delta L_{+}$ with
 
\begin{equation}
\delta L_{\pm }=-\frac{G\mu ^{2}}{2c^{2}\bar{L}^{3}}(3Gm\mu \mp 4\bar{A}
)S_{1}S_{2}\beta _{0}\ ,
\end{equation}
the radial equations (\ref{turningdef}) evaluated at the turning points have
the explicit form:
\begin{equation}
0=\frac{2E}{\mu }+\frac{2Gm}{r_{{}_{{}_{min}^{max}}}}-\frac{\bar{L}(\bar{L}
+2\delta L_{\pm })}{\mu ^{2}r_{{}_{{}_{min}^{max}}}^{2}}+\frac{
GS_{1}S_{2}\alpha _{0}}{c^{2}\mu r_{{}_{{}_{min}^{max}}}^{3}}\ .
\end{equation}
Here we have denoted $\alpha (0)=\alpha (\pi )=\alpha _{0}$. We seek for
solutions in the form $r_{{}_{{}_{min}^{max}}}=r_{\pm }+\epsilon _{\pm }$,
where $\epsilon _{\pm }$ are small corrections to the expressions of the
turning points of a Keplerian motion characterized by $\bar{L}$ and $\bar{A}
: $
 
\begin{equation}
r_{\pm }=\frac{Gm\mu \pm \bar{A}}{-2E}=\frac{\bar{L}^{2}}{\mu (Gm\mu \mp
\bar{A})}\ .
\end{equation}
The turning points are
\begin{eqnarray}
r_{{}_{{}_{min}^{max}}} &=&\frac{Gm\mu \pm \bar{A}}{-2E}-\frac{G\mu }{2c^{2}
\bar{A}\bar{L}^{2}}S_{1}S_{2}\rho _{\mp }\ ,  \nonumber \\
\rho _{\mp } &=&\alpha _{0}(\bar{A}\mp Gm\mu )+\beta _{0}(4\bar{A}\mp 3Gm\mu
)\ .  \label{turningpoints}
\end{eqnarray}
 
Then we define the generalized true anomaly parameter $\chi $ by the
relation
\begin{equation}
\frac{2}{r}=\left( \frac{1}{r_{min}}+\frac{1}{r_{max}}\right) +\left( \frac{1
}{r_{min}}-\frac{1}{r_{max}}\right) \cos \chi \ ,  \label{paramdef}
\end{equation}
which yields the expression
 
\begin{equation}
r=\frac{\bar{L}^{2}}{\mu (Gm\mu +\bar{A}\cos \chi )}-\frac{G\mu
S_{1}S_{2}\Lambda }{2c^{2}\bar{A}\bar{L}^{2}(Gm\mu +\bar{A}\cos \chi )^{2}}\
,  \label{param}
\end{equation}
with
\begin{eqnarray}
\Lambda &=&\bar{A}[\bar{A}^{2}(\alpha _{0}\!+\!4\beta_{0})
\!+\!(Gm\mu)^{2}(3\alpha _{0}\!+\!10\beta _{0})]  \nonumber \\
&+&\!Gm\mu [\bar{A}^{2}(3\alpha _{0}\!+\!11\beta_{0}) \!+\!(Gm\mu
)^{2}(\alpha _{0}\!+\!3\beta _{0})]\cos \!\chi \ .
\end{eqnarray}
 
Applications of the true anomaly parametrization (\ref{param}) will emerge
in the next section when the instantaneous expressions of radiative losses,
represented here by a generic function $f$ will be averaged as
\begin{equation}
\langle f\rangle =\frac{1}{T}\int_{0}^{2\pi }f(\chi )\frac{dt}{d\chi }d\chi
\ .  \label{ave}
\end{equation}
In $dt/d\chi =(1/\dot{r})(dr/d\chi )$ the first factor comes from a Taylor
series expansion of the radial equation (\ref{rdot2Ang}) and the second is
obtained from the derivative of Eq. (\ref{paramdef}):
\begin{equation}
\frac{dr}{d\chi }=\frac{1}{2}\left( \frac{1}{r_{min}}-\frac{1}{r_{max}}
\right) r^{2}\sin \chi \ .  \label{drdchi}
\end{equation}
The parametrization $z=\exp (i\chi )$ has the remarkable property that the
integrals (\ref{ave}) are given by the residues in the origin \cite{param}.
(The only exception under this rule is the {\it time }average{\it \ }$
\langle L\rangle $, which is treated in the Appendix.)
 
$T$ is the radial period, to be conveniently computed by means of the
generalized eccentric anomaly parametrization, defined as
\begin{equation}
2r=\left( r_{max}+r_{min}\right) -\left( r_{max}-r_{min}\right) \cos \xi \ ,
\label{xiparamdef}
\end{equation}
which yields the explicit form
 
\begin{equation}
r=\frac{Gm\mu -\bar{A}\cos \xi }{-2E}-\frac{G\mu S_{1}S_{2}\Xi }{2c^{2}\bar{
A }\bar{L}^{2}}\ ,  \label{paramxi}
\end{equation}
with
\begin{equation}
\Xi =\bar{A}(\alpha _{0}+4\beta _{0})+Gm\mu (\alpha _{0}+3\beta _{0})\cos
\!\xi \ .
\end{equation}
For the evaluation of the period
\begin{equation}
T=\int_{0}^{2\pi }\frac{dt}{d\xi }d\xi =\int_{0}^{2\pi }\frac{1}{\dot{r}}
\frac{dr}{d\xi }\ d\xi \ ,  \label{perioddef}
\end{equation}
the factor $1/\dot{r}$ is expressed again from Eq.(\ref{rdot2Ang}) by a
Taylor series expansion. However this time we have to employ also the
relations between the Newtonian true and eccentric anomaly parameters
 
\begin{equation}
\cos \chi =\frac{Gm\mu \cos \xi -\bar{A}}{Gm\mu -\bar{A}\cos \xi }\ ,\qquad
\sin \chi =\frac{\sqrt{-2E/\mu }\bar{L}\sin \xi }{Gm\mu -\bar{A}\cos \xi }\ .
\end{equation}
Either the immediate integration or the residue theorem\footnote{
The poles of the integrand (\ref{perioddef}) inside the unit circle of the
complex parameter plane $w=\exp (i\xi )$ are in the origin and at $
w_{1}=\left[ \left( Gm\mu ^{2}-\sqrt{-2\mu E\bar{L}^{2}}\right) /\left(
Gm\mu ^{2}+\sqrt{-2\mu E\bar{L}^{2}}\right) \right] ^{1/2}\ .$ The residue
at $w_{1}$ vanishes, thus the period (\ref{period}) is given by the residue
at the origin as $T=2\pi iR(w=0)$.} yields
\begin{equation}
T=2\pi Gm\left( \frac{\mu }{-2E}\right) ^{3/2}\ .  \label{period}
\end{equation}
Similarly as in the case of the spin-orbit perturbations, here the period is
given by its Keplerian expression (with $E$ characterizing the perturbed
dynamics).
 
\section{Spin-spin contributions to the instantaneous and secular losses}
 
The instantaneous radiative losses of the energy $E$ and angular momentum $
{\bf J}$ were given in terms of symmetric trace-free (STF) multipole moments
by Thorne \cite{Thorne}. To the required order they are
\begin{eqnarray}
\frac{dE}{dt} &=&-\frac{G}{5c^{5}}\left[ I^{(3)jl}I^{(3)jl}+\frac{5}{189c^{2}
}I^{(4)jlk}I^{(4)jlk}+\frac{16}{9c^{2}}J^{(3)jl}J^{(3)jl}\right] \ ,
\label{ThorneE} \\
\frac{d{\bf J}}{dt} &=&-\frac{2G}{5c^{5}}\epsilon ^{ijk}\left[
I^{(2)jl}I^{(3)kl}+\frac{5}{126c^{2}}I^{(3)jlm}I^{(4)klm}+\frac{16}{9c^{2}}
J^{(2)jl}J^{(3)kl}\right] \ ,  \label{ThorneJ}
\end{eqnarray}
where $\epsilon ^{ijk}$ is the completely antisymmetric Levy-Civita symbol,
any number in brackets denotes a corresponding order derivative and the
expressions of the symmetric trace-free (STF) multipole moments can be
derived by use of the Blanchet-Damour-Iyer formalism \cite{BDI}. The
Newtonian part of the mass quadrupole moment $I_{N}^{\ jl}$ and the
spin-orbit part of the velocity quadrupole moment $J_{SO}^{jl}$ are given by
\begin{eqnarray}
I_{N}^{\ jl} &=&\mu \left( x^{j}x^{l}\right) ^{STF}\ ,  \label{massquad} \\
J_{SO}^{\ jl} &=&\frac{3\mu }{2}\left[ x^{j}\left( \frac{{\bf S}_{{\bf 1}}}{
m_{1}}-\frac{{\bf S}_{{\bf 2}}}{m_{2}}\right) ^{l}\right] ^{STF}\ ,
\label{velquad}
\end{eqnarray}
respectively. As estimates of post-Newtonian orders show \cite{Kidder},
these are the only moments needed for evaluation of the spin-spin type
contribution to radiation losses.
 
\subsection{Energy loss}
 
The Newtonian and spin-spin contribution to the instantaneous radiative loss
in the energy is found from Eq. (\ref{ThorneE}):
\begin{equation}
\frac{dE}{dt}=-\frac{G}{5c^{5}}I_{N}^{(3)jl}({\bf a}_{N})I_{N}^{(3)kl}({\bf a
}_{N})-\frac{2G}{5c^{5}}\left[ I_{N}^{(3)jl}({\bf a}_{SS})I_{N}^{(3)kl}({\bf
a}_{N})+\frac{8}{9c^{2}}J_{SO}^{(3)jl}({\bf a}_{N})J_{SO}^{(3)kl}({\bf a}
_{N})\right] \ .  \label{ElossK}
\end{equation}
The arguments of time derivatives of the momenta contain the contribution to
the acceleration to be inserted in the respective terms. The leading order
contribution to Eq. (\ref{ElossK}) was given by Peters \cite{Peters} and the
$S_{1}S_{2}$ contribution by Kidder \cite{Kidder}, in terms of ${\bf r,v,S}_{
{\bf i}}$, and $\dot{r}$. Remarkably the last term of Eq. (\ref{ElossK})
generates also {\it self-interaction} contributions (containing either $
S_{1}^{2}$ or $S_{2}^{2})$ \cite{Kidder2}. Putting together all these terms,
the radiated energy has the expression
\begin{eqnarray}
\frac{dE}{dt} &=&-\frac{8G^{3}m^{2}\mu ^{2}}{15c^{5}r^{4}}(12v^{2}-11\dot{r}
^{2})  \nonumber \\
&&-\frac{4G^{3}m\mu }{15c^{7}r^{8}}\{-171r\dot{r}[\left( {\bf v}\cdot {\bf S}
_{{\bf 1}}\right) \left( {\bf r}\cdot {\bf S}_{{\bf 2}}\right) +\left( {\bf r
}\cdot {\bf S}_{{\bf 1}}\right) \left( {\bf v}\cdot {\bf S}_{{\bf 2}}\right)
]  \nonumber \\
&&+r^{2}[3(47v^{2}-55\dot{r}^{2})\left( {\bf S}_{{\bf 1}}\cdot {\bf S}_{{\bf
2}}\right) -3\left( 168v^{2}-269\dot{r}^{2}\right) \left( {\bf r}\cdot {\bf S
}_{{\bf 1}}\right) \left( {\bf r}\cdot {\bf S}_{{\bf 2}}\right) +71\left(
{\bf v}\cdot {\bf S}_{{\bf 1}}\right) \left( {\bf v}\cdot {\bf S}_{{\bf 2}
}\right) ]\}  \nonumber \\
&&-\frac{2G^{3}m^{2}\mu ^{2}}{15c^{7}r^{8}}\sum_{i}\frac{1}{m_{i}^{2}}
[3r^{2}(v^{2}+3\dot{r}^{2})S_{i}^{2}+9\dot{r}^{2}\left( {\bf r}\cdot {\bf S}
_{{\bf i}}\right) ^{2}-6r\dot{r}\left( {\bf r}\cdot {\bf S}_{{\bf i}}\right)
\left( {\bf v}\cdot {\bf S}_{{\bf i}}\right) +r^{2}\left( {\bf v}\cdot {\bf S
}_{{\bf i}}\right) ^{2}]\ .\   \label{ElossKidd}
\end{eqnarray}
The summation in the last term is understood over the components of the
binary.
 
This expression, however is not suitable for averaging by the method
described previously. Therefore we insert in the Newtonian terms of Eq. (\ref
{ElossKidd}) the expressions (\ref{v2Ang}) and (\ref{rdot2Ang}) for $v^{2}$
and $\dot{r}^{2}$, respectively, with $L(\chi )$ and $E(r,\chi )$ given by
Eqs. (\ref{LchiAng1}) and (\ref{ESSAng}). Then we insert in the spin-spin
terms of Eq. (\ref{ElossKidd}) the Newtonian expressions for $v^{2}$, the
Eqs. (\ref{aux1}), (\ref{aux3}) (both with $\bar{L}$ in place of $L$), Eq. (
\ref{psiN}) and
\begin{equation}
\dot{r}=\frac{\bar{A}}{\bar{L}}\sin \chi \ ,
\end{equation}
obtaining the following expression for the instantaneous loss of the energy
in terms of the true anomaly parameter $\chi $ and radius $r$ alone:
 
\begin{eqnarray}
\frac{dE}{dt} &=&\left( \frac{dE}{dt}\right) _{N}+\left( \frac{dE}{dt}
\right) _{SS-self}+\left( \frac{dE}{dt}\right) _{S_{1}S_{2}}\ ,
\label{ElossInst} \\
\left( \frac{dE}{dt}\right) _{N} &=&-\frac{8G^{3}m^{2}}{15c^{5}r^{6}}(2\mu
E\!r^{2}\!+\!2Gm\mu ^{2}r\!+\!11\bar{L}^{2})\ ,  \label{ElossInstN} \\
\left( \frac{dE}{dt}\right) _{SS-self} &=&-\frac{G^{3}m^{2}}{15c^{7}r^{8}}
\sum_{i}\left( \frac{S_{i}}{m_{i}}\right) ^{2}\{(6+sin^{2}\kappa _{i})(8\mu
E\!r^{2}\!+\!8Gm\mu ^{2}r\!-3\bar{L}^{2})  \nonumber \\
&&+sin^{2}\kappa _{i}[4\mu \bar{A}r\sin \chi \sin 2(\chi +\psi _{0}-\psi
_{i})+(8\mu E\!r^{2}\!+\!8Gm\mu ^{2}r\!-5\bar{L}^{2})\cos 2(\chi +\psi
_{0}-\psi _{i})]\}\ ,  \label{ElossInstSSself} \\
\left( \frac{dE}{dt}\right) _{S_{1}S_{2}} &=&\frac{2G^{3}mS_{1}S_{2}}{
15c^{7}\mu \bar{L}^{2}r^{8}}\left\{ \sum_{n=1}^{3}a_{n}\cos [n\chi +2(\psi
_{0}-\!\bar{\psi})]\sin \kappa _{1}\sin \kappa _{2}+a_{4}\cos \kappa
_{1}\cos \kappa _{2}+a_{5}\cos \gamma \right\} \ .  \label{ElossInstSS}
\end{eqnarray}
The coefficients $a_{k}$ are
\begin{eqnarray}
a_{1} &=&4\mu \bar{A}r(33Gm\mu ^{2}r\!-25\bar{L}^{2})\ ,  \nonumber \\
a_{2} &=&-64\mu E\bar{L}^{2}r^{2}\!+132G^{2}m^{2}\mu ^{4}r^{2}-52Gm\mu ^{2}
\bar{L}^{2}r\!+607\bar{L}^{4}\ ,  \nonumber \\
a_{3} &=&4\mu \bar{A}r(11Gm\mu ^{2}r\!+25\bar{L}^{2})\ ,  \nonumber \\
a_{4} &=&\bar{L}^{2}(64\mu Er^{2}\!+52Gm\mu ^{2}r\!-\!465\bar{L}^{2})\ ,
\nonumber \\
a_{5} &=&\bar{L}^{2}(32\mu Er^{2}\!+36Gm\mu ^{2}r\!+135\bar{L}^{2})\ .
\end{eqnarray}
By use of the true anomaly parametrization $r(\chi )$, Eq. (\ref{param}) we
find the energy loss in terms of $\chi $ alone. When we pass to the complex
parameter $z=\exp (i\chi )$ the only pole of $dE/dt$ is in the origin.
Applying the method described in the previous section, the secular change of
the energy emerges in terms of constants of motion and the average value $
\bar{L}$ of the magnitude of orbital momentum:
\begin{eqnarray}
\left\langle {\frac{dE}{dt}}\right\rangle  &=&\left\langle {\frac{dE}{dt}}
\right\rangle _{N}+\left\langle {\frac{dE}{dt}}\right\rangle
_{SS-self}+\left\langle {\frac{dE}{dt}}\right\rangle _{S_{1}S_{2}}\ ,
\label{AvelossE} \\
\left\langle {\frac{dE}{dt}}\right\rangle _{N} &=&-\frac{G^{2}m(-2E\mu
)^{3/2}}{15c^{5}\bar{L}^{7}}(148E^{2}\bar{L}^{4}+732G^{2}m^{2}\mu ^{3}E\bar{L
}+425G^{4}m^{4}\mu ^{6})\ ,  \label{AvelossEN} \\
\left\langle {\frac{dE}{dt}}\right\rangle _{SS-self} &=&\frac{G^{2}m\mu
(-2E\mu )^{3/2}}{960c^{7}\bar{L}^{11}}\sum_{i}\left( \frac{S_{i}}{m_{i}}
\right) ^{2}[C_{1}\sin ^{2}\kappa _{i}\cos 2(\psi _{0}\!-\!\psi
_{i})+C_{2}(6+\sin ^{2}\kappa _{i})]\ ,  \label{AvelossESSself} \\
\left\langle {\frac{dE}{dt}}\right\rangle _{S_{1}S_{2}} &=&-\frac{
G^{2}(-2E\mu )^{3/2}S_{1}S_{2}}{480c^{7}\bar{L}^{11}}(C_{3}\sin \kappa
_{1}\sin \kappa _{2}\cos 2(\psi _{0}\!-\!\bar{\psi})+C_{4}\cos \kappa
_{1}\cos \kappa _{2}+C_{5}\cos \gamma )\ ,  \label{AvelossESS}
\end{eqnarray}
with the following coefficients $C_{k}:$
\begin{eqnarray}
C_{1} &=&-(72E^{3}\bar{L}^{6}+428G^{2}m^{2}\mu ^{3}E^{2}\bar{L}
^{4}+406G^{4}m^{4}\mu ^{6}E\bar{L}^{2}+105G^{6}m^{6}\mu ^{9})\ ,  \nonumber
\\
C_{2} &=&-4(72E^{3}\bar{L}^{6}+660G^{2}m^{2}\mu ^{3}E^{2}\bar{L}
^{4}+910G^{4}m^{4}\mu ^{6}E\bar{L}^{2}+315G^{6}m^{6}\mu ^{9})\ ,  \nonumber
\\
C_{3} &=&-(60744E^{3}\bar{L}^{6}+602668G^{2}m^{2}\mu ^{3}E^{2}\bar{L}
^{4}+819798G^{4}m^{4}\mu ^{6}E\bar{L}^{2}+266825G^{6}m^{6}\mu ^{9})\ ,
\nonumber \\
C_{4} &=&4(17064E^{3}\bar{L}^{6}+241140G^{2}m^{2}\mu ^{3}E^{2}\bar{L}
^{4}+453670G^{4}m^{4}\mu ^{6}E\bar{L}^{2}+199899G^{6}m^{6}\mu ^{9})\ ,
\nonumber \\
C_{5} &=&-12(2056E^{3}\bar{L}^{6}+28260G^{2}m^{2}\mu ^{3}E^{2}\bar{L}
^{4}+52430G^{4}m^{4}\mu ^{6}E\bar{L}^{2}+22911G^{6}m^{6}\mu ^{9})\ .
\end{eqnarray}
 
\subsection{Loss in the magnitude of orbital angular momentum}
 
A somewhat more complicated computation yields the similar expressions for
the instantaneous and averaged radiative losses in $L.$ To complete the task
we pass through the following steps. We start from the instantaneous
radiative loss of ${\bf J}$ given in Ref. \cite{Kidder}:
\begin{eqnarray}
\frac{d{\bf J}^{i}}{dt} &=&-\frac{8G^{2}m\mu }{5c^{5}r^{3}}{\bf L}
_{N}^{i}\left( 2v^{2}-3\dot{r}^{2}+\frac{2Gm}{r}\right)   \nonumber \\
&&-\frac{2G}{5c^{5}}\epsilon ^{ijk}\left[ I_{N}^{(2)jl}({\bf a}
_{N})I_{N}^{(3)kl}({\bf a}_{SS})+I_{N}^{(2)jl}({\bf a}_{SS})I_{N}^{(3)kl}(
{\bf a}_{N})+\frac{16}{9c^{2}}J_{SO}^{(2)jl}({\bf a}_{N})J_{SO}^{(3)kl}({\bf
a}_{N})\right] \ .  \label{JlossKidd}
\end{eqnarray}
 
We compute the instantaneous radiative loss in $L$ as
\begin{equation}
\frac{dL}{dt}={\bf \hat{L}\cdot }\frac{d{\bf L}}{dt}={\bf \hat{L}\cdot }
\frac{d({\bf J-S)}}{dt}={\bf \hat{L}\cdot }\frac{d{\bf J}}{dt}-{\bf \hat{L}
\cdot }\frac{d{\bf S_{1}}}{dt}-{\bf \hat{L}\cdot }\frac{d{\bf S_{2}}}{dt}\ .
\label{Lloss}
\end{equation}
In Ref. \cite{GPV3}, starting from the Burke-Thorne potential, we have
already derived the radiative loss in the spins
\begin{equation}
\frac{1}{S_{i}}\frac{d\left( {\bf S_{i}}\right) _{\mu }}{dt}={\frac{2G}{
5c^{5}\Omega _{i}}}\left( {\frac{\Theta _{i}}{\Theta _{i}^{\prime }}}
-1\right) \epsilon _{\mu \nu \rho }I_{N}^{(5)\nu \sigma }({\bf \hat{S}_{i}}
)_{\rho }({\bf \hat{S}_{i}})_{\sigma }\ ,  \label{Sdirdot}
\end{equation}
$\Theta _{i}$ and $\Theta _{i}^{\prime }$ being the principal moments of
inertia and $\Omega _{i}$ the angular velocity of the $i$th spinning
axisymmetric body, related to the spins by $S_{i}=\Theta _{i}^{\prime
}\Omega _{i}.$ These losses appear at the second post-Newtonian order, thus
they are not negligible in our approach. From Eq. (\ref{Sdirdot}) $d{\bf S_{i}}
/dt$ are found to be perpendicular to the respective spins ${\bf S_{i}}$,
therefore $d{\bf S_{i}}/dt=$ $S_{i}d{\bf \hat{S}_{i}}/dt$. Also in Ref. \cite
{GPV3} we have shown that
\begin{equation}
\left\langle {\bf \hat{L}}\cdot \frac{d{\bf \hat{S}_{i}}}{dt}\right\rangle
=0\ .
\end{equation}
This result still holds here, in spite of the fact that the dynamics of the
orbital motion and consequently the true anomaly parametrization employed
for averaging are different. What matters in the derivation of the above
result is only the Keplerian limit of the parametrizations, which by
definition, coincide.
 
Therefore in the secular loss $\langle dL/dt\rangle $ only the first term of
Eq. (\ref{Lloss}) is relevant, which will be computed in what follows. As
PN, SO, and 2PN contributions are not considered in our treatment, we insert
${\bf L}_{N}{\bf =L}$ in the Newtonian term of (\ref{JlossKidd}), together
with the previously derived expressions (\ref{v2Ang}) and (\ref{rdot2Ang})
for $v^{2}$ and $\dot{r}^{2}$, respectively. The spin-spin perturbation term
of Eq. (\ref{JlossKidd}) can be easily dealt with by use of the Cartesian
coordinates $(x=r\cos \psi ,\ y=r\sin \psi ,\ z=0)$. Then to the required,
leading order ${\bf \hat{L}=(}0,0,1)$ (see also Fig.1. of Ref. \cite{GPV3}).
We employ in addition the Keplerian relations (\ref{dtdchiN}) and (\ref{psiN}
) which imply
\begin{equation}
\dot{\psi}=\frac{\bar{L}}{\mu r^{2}}\ .
\end{equation}
 
A lengthy but straightforward computation yields the following implicit
function of the parameter $\chi $ for ${\bf \hat{L}\cdot }d{\bf J/}dt$:
\begin{eqnarray}
{\bf \hat{L}\cdot }\frac{d{\bf J}}{dt} &=&\left( {\bf \hat{L}\cdot }\frac{d
{\bf J}}{dt}\right) _{N}+\left( {\bf \hat{L}\cdot }\frac{d{\bf J}}{dt}
\right) _{SS-self}+\left( {\bf \hat{L}\cdot }\frac{d{\bf J}}{dt}\right)
_{S_{1}S_{2}}\ ,  \label{Llossinst} \\
\left( {\bf \hat{L}\cdot }\frac{d{\bf J}}{dt}\right) _{N} &=&\frac{8G^{2}m
\bar{L}}{5c^{5}\mu r^{5}}\left( 2\mu Er^{2}-3\bar{L}^{2}\right) \ ,
\label{LlossN} \\
\left( {\bf \hat{L}\cdot }\frac{d{\bf J}}{dt}\right) _{SS-self} &=&-\frac{
2G^{3}m^{2}\mu \bar{L}}{5c^{7}r^{6}}\sum_{i}\left( \frac{S_{i}}{m_{i}}
\right) ^{2}(1+\sin ^{2}\kappa _{i})\ ,  \label{LlossSSself} \\
\left( {\bf \hat{L}\cdot }\frac{d{\bf J}}{dt}\right) _{S_{1}S_{2}} &=&\frac{
4G^{2}S_{1}S_{2}}{5c^{7}\mu ^{2}\bar{L}^{3}r^{7}}\left\{
\sum_{n=1}^{3}b_{n}\cos [n\chi +2(\psi _{0}-\!\bar{\psi})]\sin \kappa
_{1}\sin \kappa _{2}+b_{4}\cos \kappa _{1}\cos \kappa _{2}+b_{5}\cos \gamma
\right\} \ ,  \label{LlossSS}
\end{eqnarray}
with the coefficients $b_{k}$ given by
\begin{eqnarray}
b_{1} &=&3\mu \bar{A}r(-2Gm\mu ^{3}Er^{3}+\mu E\bar{L}^{2}r^{2}+9Gm\mu ^{2}
\bar{L}^{2}r-8\bar{L}^{4})\ ,  \nonumber \\
b_{2} &=&3(-2G^{2}m^{2}\mu ^{5}Er^{4}-22\mu E\bar{L}^{4}r^{2}+9G^{2}m^{2}\mu
^{4}\bar{L}^{2}r^{2}-10Gm\mu ^{2}\bar{L}^{4}r+19\bar{L}^{6})\ ,  \nonumber \\
b_{3} &=&\mu \bar{A}r(-2Gm\mu ^{3}Er^{3}-3\mu E\bar{L}^{2}r^{2}+9Gm\mu ^{2}
\bar{L}^{2}r+24\bar{L}^{4})\ ,  \nonumber \\
b_{4} &=&\bar{L}^{4}(54\mu Er^{2}+23Gm\mu ^{2}r-45\bar{L}^{2})\ ,  \nonumber
\\
b_{5} &=&-3\bar{L}^{4}(6\mu Er^{2}+2Gm\mu ^{2}r-5\bar{L}^{2})\ .
\end{eqnarray}
We would like to stress the appearance of the {\it self-interaction terms} (
\ref{LlossSSself}) in the instantaneous loss of $L$ either. They originate
in the last term of Eq. (\ref{JlossKidd}).
 
A similar averaging procedure as in the case of the energy loss, gives here
\begin{eqnarray}
\left\langle {\frac{dL}{dt}}\right\rangle  &=&\left\langle {\frac{dL}{dt}}
\right\rangle _{N}+\left\langle {\frac{dL}{dt}}\right\rangle
_{SS-self}+\left\langle {\frac{dL}{dt}}\right\rangle _{S_{1}S_{2}}\ ,
\label{avelossL} \\
\left\langle {\frac{dL}{dt}}\right\rangle _{N} &=&-\frac{4G^{2}m(-2\mu
E)^{3/2}}{5c^{5}\bar{L}^{4}}(14E\bar{L}^{2}+15G^{2}m^{2}\mu ^{3})\ ,
\label{avelossLN} \\
\left\langle {\frac{dL}{dt}}\right\rangle _{SS-self} &=&-\frac{G^{2}m\mu
(-2\mu E)^{3/2}}{20c^{7}\bar{L}^{8}}D_{1}\sum_{i}\left( \frac{S_{i}}{m_{i}}
\right) ^{2}(1+\sin ^{2}\kappa _{i})\ ,  \label{avelossLSSself} \\
\left\langle {\frac{dL}{dt}}\right\rangle _{S_{1}S_{2}} &=&\frac{G^{2}(-2\mu
E)^{3/2}S_{1}S_{2}}{10c^{7}\bar{L}^{8}}[D_{2}\sin \kappa _{1}\sin \kappa
_{2}\cos 2(\psi _{0}\!-\!\bar{\psi})+D_{3}\cos \kappa _{1}\cos \kappa
_{2}+D_{4}\cos \gamma ]\ ,  \label{avelossLSS}
\end{eqnarray}
where the coefficients $D_{k}$ are given below
\begin{eqnarray}
D_{1} &=&12E^{2}\bar{L}^{4}+60G^{2}m^{2}\mu ^{3}E\bar{L}^{2}+35G^{4}m^{4}\mu
^{6}\ ,  \nonumber \\
D_{2} &=&12(62E^{2}\bar{L}^{4}+211G^{2}m^{2}\mu ^{3}E\bar{L}
^{2}+90G^{4}m^{4}\mu ^{6})\ ,  \nonumber \\
D_{3} &=&-(1524E^{2}\bar{L}^{4}+7260G^{2}m^{2}\mu ^{3}E\bar{L}
^{2}+4865G^{4}m^{4}\mu ^{6})\ ,  \nonumber \\
D_{4} &=&24(22E^{2}\bar{L}^{4}+105G^{2}m^{2}\mu ^{3}E\bar{L}
^{2}+70G^{4}m^{4}\mu ^{6})\ .
\end{eqnarray}
We emphasize that self-interaction terms (\ref{AvelossESSself}) and (\ref
{avelossLSSself}) are present in both the secular losses of $E$ and $L$.
 
\section{Concluding remarks}
 
The main result of this paper consists in the derivation of the the second
post-Newtonian order spin-spin contribution to the radiation back reaction
on the energy and magnitude of orbital angular momentum of the coalescing
binary system. In order to complete the task, a solution $r(\chi )$ of the
quasi-Keplerian radial motion of the binary, with the spin-spin interaction
as the perturbation, was derived in terms of the true anomaly
parametrization. The secular losses of $E$, Eqs. (\ref{AvelossE})-(\ref
{AvelossESS}) and of $L$, Eqs. (\ref{avelossL})-(\ref{avelossLSS}), in
complement with the corresponding spin-orbit terms from Eqs. (4.1), (4.2) of
Ref. \cite{GPV3} (with the replacements $L\rightarrow \bar{L}$ and $
A_{0}\rightarrow \bar{A}$, which are allowed up to the $3/2$ post-Newtonian
order), represent the total spin-induced back reaction of the radiation up
to the second post-Newtonian order. Among the loss terms due to the
spin-spin interaction, terms originating in the self-interaction of the
spins were found. While all other spin-spin terms vanish in the one-spin
limit ${\bf S_{2}}=0$, (half of) these terms survive, representing the
second post-Newtonian correction to the losses derived earlier in the
Lense-Thirring approximation \cite{GPV1}.
 
In the nonradiative description of the angles $\kappa _{i}$ and $\gamma $
in principle not all second post-Newtonian order spin terms can be found
from the Lagrangian (\ref{LagSS}). Similarly as the leading spin-orbit and
spin-spin terms in the spin precession equations (exceptionally) are of
first and $3/2$ post-Newtonian orders, respectively \cite{GPV3}, any higher
order description of the evolution of the spinning binary then Eq. (\ref{LagSS})
could give new contributions even at the second post-Newtonian order. Such a
description is in the process of completion \cite{Owen}. On the other hand,
the situation is more sophisticated when it comes to the radiative evolution
of the angles $\kappa _{i}$ and $\gamma $. New terms beyond those already
given in Ref. \cite{GPV3} can arise in two ways. The first type of these
terms would come from corrections to the Burke-Thorne potential \cite{BT}.
It was already shown \cite{Blanchet} that the first correction is one
post-Newtonian order higher than the potential itself, therefore any
correction from the radiative loss of the spins would contribute only at $5/2
$ post-Newtonian order (above the leading radiation term in the orbital
angular momentum loss). However new terms in the radiative angular evolution
appear at the second post-Newtonian order due to $(d{\bf J}/dt)_{SS}$. A
computation in progress reveals a complicated structure of these terms. We
defer the study of the angular evolutions at the second post-Newtonian order
to a subsequent work.
 
\section{Acknowledgments}
 
The author is grateful to Lawrence E. Kidder for bringing into his attention
the self-interaction spin terms in the energy loss. This work has been
supported by the Soros Foundation and the Hungarian Scholarship Board. The
algebraic package REDUCE was employed in some of the computations.
 
\appendix
 
\section{The relation between time and angular averages of $L$}
 
We start from the expression (\ref{LchiAng1}) for $L(\chi )$. The time
average is obtained as
\begin{equation}
\langle L\rangle =\frac{1}{T}\int_{0}^{T}L(\chi )dt=\frac{1}{T}
\int_{0}^{2\pi }L(\chi )\frac{dt}{d\chi }d\chi \ .  \label{Lavetime}
\end{equation}
Here $T$ is given by Eq. (\ref{period}) and $dt/d\chi $ is derived from the
radial equation as explained in Sec. III. When expressed in terms of the
complex parameter $z=\exp (i\chi )$, the integrand in Eq. (\ref{Lavetime})
has two poles inside the unit circle: one in the origin $z_{0}=0$ and the
other at
\begin{equation}
z_{1}=-w_{1}=-\left[ \frac{Gm\mu ^{2}-\sqrt{-2\mu E\bar{L}^{2}}}{Gm\mu ^{2}+
\sqrt{-2\mu E\bar{L}^{2}}}\right] ^{1/2}\ .
\end{equation}
From the residue theorem, $\langle L\rangle $ is given as the sum of the
residues at $z_{0}$ and $z_{1}$ multiplied with $2\pi i/T$.
 
Thus the desired relation between the angular average $\bar{L}$ and the time
average $\langle L\rangle $ emerges as
\begin{equation}
\langle L\rangle =\bar{L}-\frac{G\mu E_{1}}{2c^{2}\bar{A}^{2}\bar{L}
^{3}E_{2} }S_{1}S_{2}\sin \kappa _{1}\sin \kappa _{2}\cos 2(\psi _{0}\!-\!
\bar{\psi})  \label{LaveLbar}
\end{equation}
with the coefficients $E_{1,2}$ given by
\begin{eqnarray}
E_{1} &=&2(-2\mu E)^{1/2}\bar{L}[\bar{A}^{6}-15G^{2}m^{2}\mu ^{2}\bar{A}
^{4}+32G^{4}m^{4}\mu ^{4}\bar{A}^{2}-16G^{6}m^{6}\mu ^{6}]  \nonumber \\
&&+Gm\mu ^{2}[-11\bar{A}^{6}+58G^{2}m^{2}\mu ^{2}\bar{A}^{4}-80G^{4}m^{4}\mu
^{4}\bar{A}^{2}+32G^{6}m^{6}\mu ^{6}]  \nonumber \\
E_{2} &=&4Gm(-2\mu E)^{1/2}\bar{L}[\bar{A}^{2}-2G^{2}m^{2}\mu ^{2}]+\bar{A}
^{4}-8G^{2}m^{2}\mu ^{2}\bar{A}^{2}+8G^{4}m^{4}\mu ^{4}\ .
\end{eqnarray}
Whenever needed, all expressions can be rewritten in terms of $\langle
L\rangle $ by use of Eq. (\ref{LaveLbar}). However, by inserting $\bar{L}$
expressed from Eq. (\ref{LaveLbar}) in Eq. (\ref{LchiAng1}) we obtain a
considerably longer expression for $L(\chi )$ in terms of $\langle L\rangle $
than the original Eq. (\ref{LchiAng1}). This is an indication that the
description in terms of the angular average $\bar{L}$ is better suited for
the problem.

\end{document}